\documentstyle[subeqn,preprint,aps,epsfig]{revtex}
\begin{document}
 
\def\real{I\negthinspace R}
\def\zed{Z\hskip -3mm Z }
\def\half{\textstyle{1\over2}}
\def\quarter{\textstyle{1\over4}}
\newcommand{\be}{\begin{equation}}
\newcommand{\ee}{\end{equation}}
\preprint{DTP/97/43, gr-qc/9709029}
\draft
\tighten
\renewcommand{\topfraction}{0.8}
%%%%%%%%%%%%%%%%% HERE comment out next 2 lines
%@twocolumnfalse\endcsname
 
\title{GLOBAL MONOPOLES IN DILATON GRAVITY}
\author{Owen Dando and Ruth Gregory}
\address{\  
Centre for Particle Theory, 
Durham University, South Road, Durham, DH1 3LE, U.K.
}
\date{\today}
\maketitle
\begin{abstract}
We analyse the gravitational field of a global monopole within 
the context of low energy string gravity, allowing for an
arbitrary coupling of the monopole fields to the dilaton. 
Both massive and massless dilatons are considered. We find
that, for a massless dilaton, the spacetime is generically
singular, whereas when the dilaton is massive, the monopole
generically induces a long range dilaton cloud. We compare
and contrast these results with the literature.
 
\ 

\end{abstract}

\pacs{PACS numbers: 04.70.-b, 11.27.+d, 98.80.Cq \hspace*{2cm} 
gr-qc/9709029}
 
%%%%%%%%%%%%%%%%% HERE comment out next  line
%\vskip2pc]
 
%%%%%%%%%%%%%%%%%%%%%%%%%%%%%%%%%%%%%%%%%%%%%%%%%%%%%%%%%%%%%%%%%%%%%% 

\section{Introduction}

Cosmologists have been attracted to topological defects as a 
possible source for the density perturbations which seeded 
galaxy formation\cite{BSV}. Phase transitions in the early universe 
can give rise to various types of defect. Briefly, a defect is a 
discontinuity in the vacuum, and can be classified according 
to the topology of the vacuum manifold of the field theory 
model being used. Disconnected vacuum manifolds give domain 
walls, non-simply connected manifolds, strings, 
and vacuum manifolds with non-trivial $\Pi_2$ and $\Pi_3$ homotopy groups 
give monopoles and textures respectively. Strings and monopoles can be further 
subdivided into local and global defects depending on whether the 
symmetry broken is local or global. With the exception of the domain wall,
which results from the breaking of a discrete symmetry and has no Goldstone 
boson, global defects are usually characterised
by a power law fall-off in the energy density of the defect leading
to divergent energies. Local defects on the other hand 
typically have no Goldstone bosons
and are characterised by a well-defined core and finite energy per
unit defect area.
We therefore expect local and global defects to have significantly 
different behaviour, and nowhere is this more evident than 
in their coupling to gravity. Whilst local strings \cite{LSTR} and 
monopoles\cite{LMON} produce only localised spacetime curvature, asymptoting
locally flat and flat spacetimes respectively, global strings\cite{GSTR,G} 
and monopoles\cite{BV,HL} have strong 
effects even at large distances. Indeed, the spacetime of both domain 
walls\cite{DW} and global strings\cite{G} appears to be time
dependent, with a de-Sitter expansion along the spatial extent of the
defect. The spacetime of a global 
monopole is static, non-singular but not asymptotically flat; 
it asymptotes a locally flat spacetime, with
a deficit solid angle of $8 \pi G \eta^{2}$ 
where $\eta$ is the symmetry breaking scale \cite{BV}, but this
deficit angle can lead to potentially strong tidal forces \cite{H}. 

Of course, this discussion has taken place within the context of
general relativity, however, at sufficiently high energy scales 
it seems likely that gravity is not given by the 
Einstein action. The most promising alternative seems to be 
that given by string theory, where in the low energy limit
gravity becomes scalar-tensor 
in nature\cite{LESG}. Scalar-tensor gravity is not new, it was
pioneered by Jordan, Brans and Dicke\cite{JBD}, 
who sought to incorporate Mach's
principle into gravity, and indeed the scalar-tensor part of the low
energy superstring action is equivalent to Brans-Dicke 
theory for a particular value of the
Brans-Dicke parameter: $\omega = -1$. 
The implications of superstrings for cosmology
is a subject of intense investigation, however, in this paper we are 
interested in the implications of superstring gravity for
topological defects, in particular global monopoles, and how these
effects are dependent on the mass of the dilaton. 

Recently, Damour and Vilenkin\cite{DV} argued that a
low mass superstring dilaton would be incompatible with a local string
network formed at a GUT phase transition, however, by considering
the fully coupled nonlinear field equations of a particular local string
model with dilaton gravity, one can show that this conclusion is
strongly dependent on the coupling of the defect to the dilaton\cite{GS}.
Here we consider the gravi-dilaton 
field of the global monopole in superstring gravity. We 
consider a general form for the interaction with the dilaton, 
assuming, as in \cite{GS}, that the monopole lagrangian 
couples to the dilaton via 
an arbitrary coupling $e^{2a\phi} \mathcal{L}$ in the string frame. 
We consider both massive and massless dilatons, which unsurprisingly
turn out to be qualitatively rather different.

The layout of the paper is as follows. We first review the work of
Barriola and Vilenkin, deriving the Einstein metric of a global monopole.
We then present the analysis for the global monopole in superstring gravity,
for both massless and massive dilatons. Finally, we consider the 
implications of the dilaton for cosmological bounds on the monopole.

\section{The global monopole}

\begin{figure}
\begin{center}
\epsfig{figure=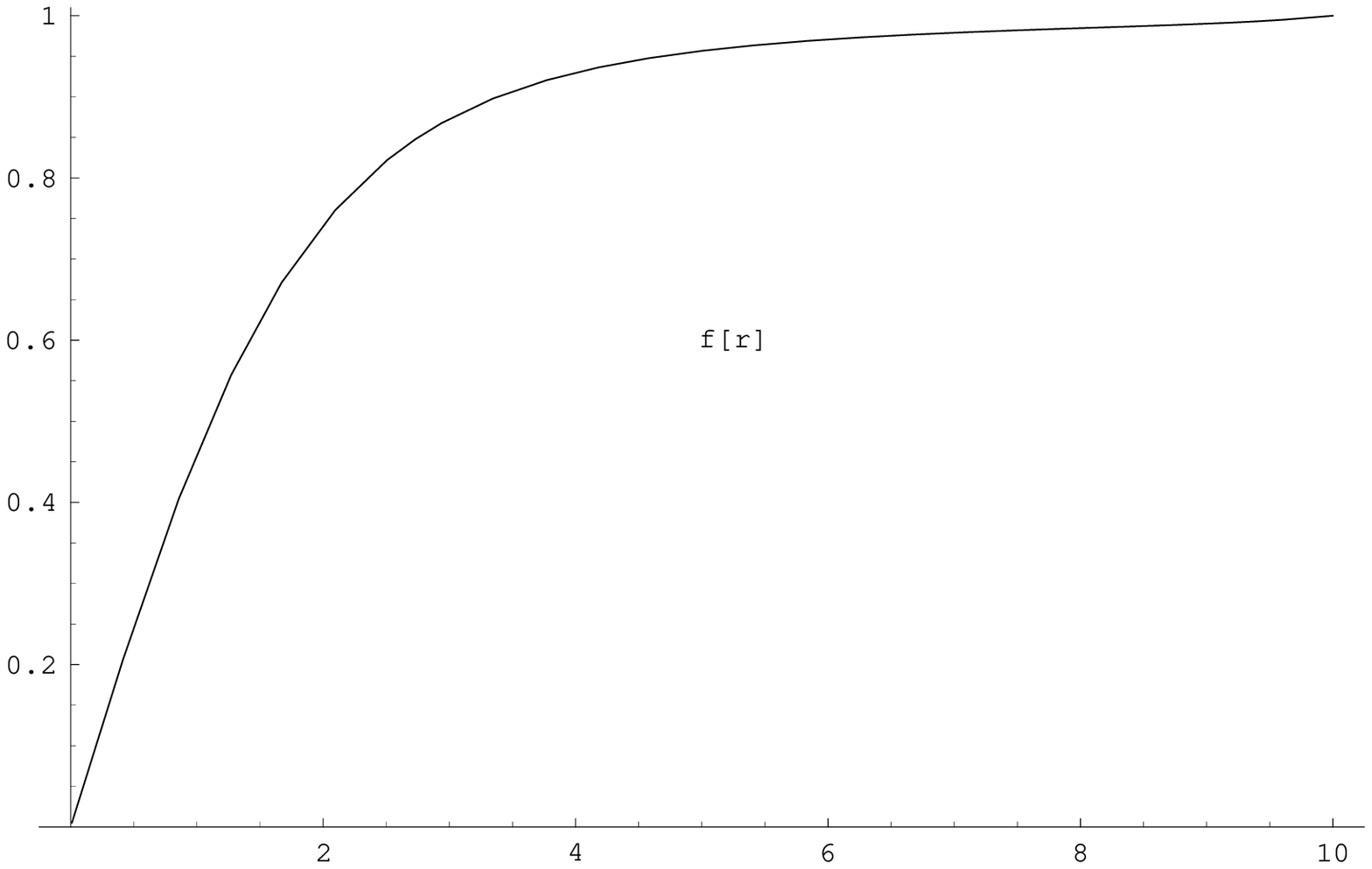,height=18cm}
\caption{The monopole field $f(r)$.
\label{fig:fofr}}
\end{center}
\end{figure}

In this section we briefly review the work of Barriola and Vilenkin \cite{BV}.
The simplest model that gives global monopoles is described by the Lagrangian
\be
{\mathcal{L}}(\psi^{i})=\frac{1}{2} \nabla_a \psi^{i} \nabla^{a} \psi^{i} 
- \frac{\lambda}{4} (\psi^{i} \psi^{i} - \eta^{2})^{2} \label{lagrangian}
\ee
where $\psi^{i}$ is a triplet of real scalar fields, ${i=1,2,3}$. 
This model has a global O(3) symmetry, which is spontaneously broken
to a global U(1) symmetry by a choice of vacuum $|\psi^i|=\eta$.
We look for a spherically symmetric, static configuration describing the global
monopole at rest. The field configuration describing a monopole may then
be written as
\be\label{fc}
\psi^{i}=\eta f(r) {\hat x}^{i}
\ee
where ${\hat x}^{i}$ is the unit radial vector in the internal space. 
The metric for the static monopole is then written as
\be\label{metric}
ds^{2}=B(r) dt^2 - A(r) dr^2 - r^2 (d\theta^2+\sin^2 \theta d\phi^2)
\ee
In terms of the new variable $f(r)$ the Lagrangian becomes
\be
{\mathcal{L}}=-\eta^2 \left (\frac{f'^2}{2A}+\frac{f^2}{r^2}
+\frac{\lambda \eta^2}{4} (f^2-1)^2 \right ) 
\ee
and the field equation for $f$ is
\be
\frac{1}{(AB)^{1/2}r^2} \left ( \left ( \frac{B}{A} \right )^{1/2} r^2 f' 
\right )' = \frac{2f}{r^2} + \lambda \eta^2 f (f^2-1)
\ee
Even in a flat space background ($A=B=1$), this equation does not have an
analytic solution, however, it can be integrated numerically (see figure 
\ref{fig:fofr}). Note that $f(r) \sim 1 - 1/r^2$ as $r\to\infty$.

The monopole couples to the metric via its energy momentum tensor
\be
G_{ab}=8\pi G T_{ab}
\ee
Without loss of generality, we will assume that the scale of the
coordinates in (\ref{metric}) is such  
that the size of the monopole core is of order 
unity, i.e.\ we set $\sqrt{\lambda} \eta =1$. We also rescale the
energy-momentum tensor of the monopole, $\hat{T}_{ab}=T_{ab}/ \eta^2$, 
which is given by
\begin{eqnarray}
& & \hat{T}_t^t= \frac{f'^2}{2A}+\frac{f^2}{r^2}+ 
\frac{1}{4} (f^2-1)^2  \nonumber \\
& & \hat{T}_r^r= - \frac{f'^2}{2A} + \frac{f^2}{r^2} + \frac{1}{4} (f^2-1)^2  \\
& & \hat{T}_{\theta}^{\theta} = \frac{f'^2}{2A} 
+ \frac{1}{4} (f^2-1)^2  \nonumber 
\end{eqnarray}
The $tt$ and $rr$ components of Einstein's equation are
\begin{eqnarray}
& & \frac{A'}{r A^2}+ \frac{1}{r^2} \left ( 1 - \frac{1}{A} \right ) 
=  \epsilon    \hat{T}_t^t \label{aeqn} \\
& & \frac{B'}{ABr} - \frac{1}{r^2} \left ( 1-\frac{1}{A} \right ) 
=  -\epsilon \hat{T}_r^r \label{beqn}  
\end{eqnarray}
where $\epsilon=8 \pi G \eta^2$ is the gravitational strength of the monopole.

Outside the monopole core, $f(r) \approx 1$ and the energy-momentum 
tensor can be approximated by
\be
\hat{T}_t^t \approx \hat{T}_r^r \approx \frac{1}{r^2}, 
\hat{T}_{\theta}^{\theta} \approx 0
\ee
The general solution to Einstein's equations is
\be
B=A^{-1}=1- \epsilon -\frac{2GM}{r}
\ee
where $M$ is a constant of integration, the ADM mass of the monopole
(see \cite{NS} for a rigorous definition of ADM mass in quasi-asymptotically
flat spacetimes), which from (\ref{aeqn}) is
\be
2GM = \epsilon \int_0^\infty \left [ r^2 {\hat T}^t_t - 1 \right ] dr
\label{gmass}
\ee
We obtain $GM = -0.73\epsilon$ in a
linearised approximation, in agreement with \cite{HL}.
Ignoring the mass term, which is negligible on astrophysical 
scales, and rescaling the $r$ and $t$ variables we can write 
the monopole metric as
\be
ds^2=dt^2-dr^2-(1-\epsilon)r^2(d\theta^2+\sin^2 \theta d\phi^2)
\ee
This metric describes a spacetime with deficit solid angle $\epsilon$. 
The spacetime is not asymptotically flat, but it is asymptotically 
locally flat.
Note that the mass is negative, which at first seems suprising since
the usual ADM mass for asymptotically flat spacetimes is necessarily
positive \cite{PMASS}. However, as discussed by Nucamendi and
Sudarsky \cite{NS}, there is no such positivity requirement for
quasi-asymptotically flat spacetimes, and indeed the negativity
of the global monopole mass is in keeping with the gravitationally
repulsive domain wall \cite{DW}, which is also a global defect.

\section{Global monopoles in dilaton gravity}

We are interested in the behaviour of the global monopole 
metric when gravitational interactions take a form typical of low 
energy string theory. In its minimal form, string gravity replaces 
the gravitational constant, $G$, by a scalar field, the dilaton. 
To account for the unknown coupling of the dilaton to the monopole, 
as in \cite{GS} we choose the action
\be
S=\int d^4 x \sqrt{-\hat{g}} \left [ e^{-2\phi} \left ( -\hat{R} 
- 4 (\hat{\nabla} \phi)^2 - \hat{V}(\phi) \right ) + e^{2a\phi} 
{\mathcal{L}} \right ] \label{stringaction}
\ee
where $\mathcal{L}$ is as in (\ref{lagrangian}). The potential for the
dilaton ${\hat V}(\phi)$ is for the moment assumed general.
This action is written in 
terms of the string metric which appears in the string sigma model. 
To facilitate comparison with the previous section we instead
choose to write the action in terms of the `Einstein' metric
\be
g_{ab}=e^{-2\phi} \hat{g}_{ab}
\ee
in which the gravitational part of the action appears in the normal Einstein 
form
\be
S=\int d^4x \sqrt{-g} \left [ -R + 2(\nabla \phi)^2 - V(\phi) 
+ e^{2(a+2)\phi} {\mathcal{L}}(\psi^{i},e^{2\phi}g) \right ]
\ee
where $V(\phi)=e^{2\phi}\hat{V}(\phi)$. The energy-momentum tensor is now
\be
T_{ab}=2 \frac{\delta {\mathcal{L}}(\psi^{i},e^{2\phi}g)}
{\delta g^{ab}} - g_{ab}{\mathcal{L}}(\psi^{i},e^{2\phi}g)= 
e^{-2\phi} \nabla_a \psi^i \nabla_b \psi^i - g_{ab}{\mathcal{L}}
\ee
Einstein's equation becomes
\be
G_{ab}=  \frac{1}{2} e^{2(a+2)\phi} T_{ab} + S_{ab}
\ee
where
\be
S_{ab}=2 \nabla_a \phi \nabla_b \phi +\frac{1}{2} g_{ab} V(\phi) 
- g_{ab} (\nabla \phi)^2
\ee
is the energy-momentum of the dilaton, which has as its
equation of motion
\be
- \Box \phi = \frac{1}{4} \frac{\partial V}{\partial \phi} 
- \frac{a+2}{2} e^{2(a+2)\phi} {\mathcal{L}}(\psi^{i},e^{2\phi}g)
+\frac{1}{4} e^{2(a+1)\phi} g^{ab} \nabla_a \psi^i \nabla_b \psi^i
\ee

As before, we choose the general static, spherically symmetric metric
(\ref{metric})
and write the field configuration (\ref{fc}) for the monopole.
The Lagrangian now is
\be
{\mathcal{L}}=-\eta^2 \left (e^{-2\phi} \left ( \frac{f'^2}{2A}
+\frac{f^2}{r^2} \right ) + \frac{\lambda \eta^2}{4} (f^2-1)^2 \right ) 
\ee
Again we take $\sqrt{\lambda} \eta =1$ and the rescaled modified 
energy-momentum tensor is then
\begin{eqnarray}
& & \hat{T}_t^t = e^{2(a+2)\phi} \left (e^{-2\phi}\left(\frac{f'^2}{2A}
+\frac{f^2}{r^2}\right)+ \frac{1}{4} (f^2-1)^2 \right )  \nonumber  \\
& & \hat{T}_r^r =  e^{2(a+2)\phi} \left (e^{-2\phi} \left(- \frac{f'^2}{2A} 
+ \frac{f^2}{r^2}\right) + \frac{1}{4} (f^2-1)^2 \right ) \label{em} \\
& & \hat{T}_{\theta}^{\theta} =  e^{2(a+2)\phi} \left (e^{-2\phi} 
\frac{f'^2}{2A} + \frac{1}{4} (f^2-1)^2 \right )  \nonumber
\end{eqnarray}
The $tt$ and $rr$ components of Einstein's equation are now
\begin{eqnarray}
& & \frac{A'}{r A^2}+ \frac{1}{r^2} \left ( 1 - \frac{1}{A} \right ) 
= \epsilon \hat{T}_t^t + \frac{1}{2} V(\phi) + \frac{\phi '^2}{A}  
\label{einstein1}\\
& & -\frac{B'}{ABr} + \frac{1}{r^2} \left ( 1-\frac{1}{A} \right ) 
= \epsilon \hat{T}_r^r + \frac{1}{2} V(\phi) -  \frac{\phi '^2}{A} 
\label{einstein2}
\end{eqnarray}
where $\epsilon = \eta^{2}/2$. 
The dilaton equation is
\be
-\Box{\phi} = \frac{1}{4} \frac{\partial V }{\partial \phi} 
+ \epsilon (a+1){\hat T}^t_t + {\epsilon\over4} e^{2(a+2)\phi}(f^2-1)^2
\label{dilaton}
\ee
and the equation of motion for $f$ is 
\be
\frac{1}{(AB)^{1/2}r^2} \left ( \left ( \frac{B}{A} \right ) ^{1/2} 
r^2 e^{2(a+1)\phi} f' \right ) ' = \frac{2f e^{2(a+1)\phi}}{r^2}+ 
e^{2(a+2)\phi} f(f^2-1) \label{feqn}
\ee
We now consider massless and massive dilatons in turn.

\subsection{Massless dilatonic gravity}

For the massless dilaton $V(\phi)=0$. Hence the dilaton equation is
\begin{eqnarray}
&  - \Box \phi & =\left ( \frac{B'}{2AB} - \frac{A'}{2A} + \frac{2}{Ar} 
\right ) \phi ' + \frac{\phi ''}{A} \nonumber \\
&  & = \epsilon \left ((a+1)e^{2(a+1)\phi} \left ( \frac{f'^2}{2A} 
+ \frac{f^2}{r^2} \right ) + \frac{a+2}{4} e^{2(a+2)\phi} (f^2-1)^2 \right ) 
\label{dilaton2} 
\end{eqnarray}
As a preliminary step, we consider linearizing the equations of
motion, expanding the functions in powers of the small parameter $\epsilon$
\begin{eqnarray}
& & A=1+\epsilon A_1 + \ldots  \nonumber \\
& & B=1+\epsilon B_1 + \ldots \\
& & \phi =\phi_0 + \epsilon \phi_1 + \ldots  \nonumber
\end{eqnarray}
To ${\mathcal{O}}(1)$, (\ref{einstein1}) gives
\be
\phi_0'^{2} = 0 
\ee
Hence $\phi_0=const$. Then to ${\mathcal{O}}(\epsilon)$ we have
\begin{eqnarray}
\frac{A_1'}{r} + \frac{A_1}{r^2} & = & e^{2(a+1)\phi_0} \left ( 
\frac{f'^2}{2}+ \frac{f^2}{r^2} \right )+ \frac{1}{4}e^{2(a+2)\phi_0}
(f^2-1)^2 \nonumber \\
-\frac{B_1'}{r} + \frac{A_1}{r^2} & = & e^{2(a+1)\phi_0} \left ( - 
\frac{f'^2}{2}+ \frac{f^2}{r^2} \right )+ \frac{1}{4}e^{2(a+2)\phi_0}(f^2-1)^2\\
\phi_1 '' + \frac{2\phi_1 '}{r} &=& (a+1) e^{2(a+1)\phi_0} 
\left ( \frac{f'^2}{2} + \frac{f^2}{r^2} \right ) 
+ \frac{a+2}{4} e^{2(a+1)\phi_0} (f^2-1)^2 \nonumber
\end{eqnarray}
For large $r$, $f \approx 1$ and 
\begin{eqnarray}
(A_1 r)' & = & e^{2(a+1)\phi_0} \nonumber \\
B_1 & = & -A_1 \\
\phi_1 '' + \frac{2\phi_1'}{r} & = & \frac{(a+1)e^{2(a+1)\phi_0}}{r^2}
\nonumber \end{eqnarray}
Then
\begin{eqnarray}
A & = & 1 + \epsilon \left ( e^{2(a+1)\phi_0} 
+ \frac{a_1}{r} \right ) + \ldots \nonumber \\
B & = & 1 - \epsilon \left ( e^{2(a+1)\phi_0} 
+ \frac{a_1}{r} \right ) + \ldots \label{linear} \\
\phi & = & \phi_0 + \epsilon \left ( c_1 + 
\frac{c_2}{r} + (a+1)e^{2(a+1)\phi_0} \ln r \right ) \nonumber
\end{eqnarray}
This agrees with the linearised result of Barros and Romero 
\cite {BR} (see also Banerjee et.\ al.\ \cite{BBCS}) who studied 
global monopoles in Brans-Dicke gravity, however, as we will show,
it is not enough to find a linearised solution, one must also
consider self-consistency of the approximation one is using. 

It appears from (\ref{linear}) that we may have 
an asymptotically locally flat spacetime, at least in the Einstein frame,
however, note that $\phi_1$ 
is divergent unless $a=-1$. Hence, unless $a = -1$, the linearized 
approximation ceases to be valid for $r\simeq e^{1/\epsilon}$ and we must
therefore consider the back reaction of the dilaton on the spacetime.
In fact by studying the full field equations, 
we can show that if $a \ne -1$ no such asymptotically locally 
flat spacetime exists for the global monopole in massless dilatonic gravity.

First note that from (\ref{einstein1}) and (\ref{einstein2}) we have 
\begin{eqnarray}
\left [ r(1-A^{-1}) \right ] ' & = & \epsilon r^2 \left ( 
e^{2(a+1)\phi} \left ( \frac{f'^2}{2A} + \frac{f^2}{r^2} 
\right ) + \frac{1}{4} e^{2(a+2)\phi}(f^2-1)^2 \right ) 
+ \frac{\phi'^{2} r^2}{A} \label{einstein1m} \\
\left [ \ln (AB) \right ]' & =  & \epsilon e^{2(a+2)\phi} r f'^2 
+ 2r\phi'^2 \label{einstein2m}
\end{eqnarray}
For a locally asymptotically flat spacetime we want $A,B 
\rightarrow const$ as $r \rightarrow \infty$. Hence by 
integrating (\ref{einstein2m}) between zero and infinity we 
see that the integrals 
\begin{eqnarray}
I_1 & = & \int_0^{\infty} e^{2(a+2)\phi} r f'^2 dr \nonumber \\
I_2 & = & \int_0^{\infty} r \phi'^{2} dr
\end{eqnarray}
must be convergent. 
Now consider $G_0^0-G_r^r-2G_{\theta}^{\theta} = 2R_0^0$: 
\be
\left ( \frac{r^2 (\sqrt{B})'}{\sqrt{A}} \right )' = -\frac{1}{4} 
\epsilon \sqrt{AB} r^2 e^{2(a+2)\phi} (f^2-1)^2
\ee
Since $A,B\to const.$, asymptotically this gives
\be
\sqrt{AB}r^2 e^{2(a+2)\phi} (f^2-1)^2 = o(1)
\label{ord}
\ee 
as $r\rightarrow \infty$. Since $A \rightarrow const \ne 1$ as 
$r\rightarrow \infty$, (\ref{einstein1m}) gives
\be
\epsilon e^{2(a+1)\phi} \frac{f'^{2}r^2}{2A} 
+ \epsilon e^{2(a+1)\phi} f^2 + \frac{1}{4} 
\epsilon e^{2(a+2)\phi} r^2 (f^2-1)^2 + 
\frac{\phi'^2 r^2}{A} \sim \kappa_1 \label{blah}
\ee
at infinity, where $\kappa_1$ is a constant. But the convergence of 
the integrals $I_i$ implies the integrands are $o(1/r)$ as 
$r\rightarrow \infty$. Together with (\ref{ord})
this means all but one term of the left hand side of 
(\ref{blah}) disappears at infinity. That is
\be
\epsilon e^{2(a+1)\phi} f^2 \sim \kappa_1
\ee 
as $r \rightarrow \infty$.

Now consider the dilaton equation (\ref{dilaton2}). Near infinity we have
\be
\frac{1}{\sqrt{AB}} \left ( \sqrt{\frac{B}{A}} 
r^2 \phi'\right )' \sim \epsilon (a+1) e^{2(a+1)\phi} f^2 \sim \kappa_2
\ee
since $a \ne -1$. That is 
\be
\phi' \sim \frac{A \kappa_2}{r}
\ee
Hence
\be
r^2 \phi'^{2} \sim A^2 \kappa_2^2
\ee
But from above, the convergence of $I_2$ implies $r^2 \phi'^{2}
\rightarrow 0$ as $r\rightarrow \infty$. Hence we have a 
contradiction, and no non-singular, locally asymptotically flat 
spacetime exists for the monopole. What has happened is that the 
energy density depends on $e^{2(a+1)\phi}$, which in turn is driven
by the energy density, thus causing a disastrous feedback effect.

Now consider the case $a=-1$. The dilaton equation is now 
\be
\left ( \sqrt{B\over A} r^2 \phi'\right )' =
{\epsilon\over 4} r^2  \sqrt{AB} e^{2\phi} (f^2-1)^2
\ee
which, in the linearized approximation, can be integrated to give
\be
\phi_1 = -{1\over 4r} \int_0^r \xi^2(f^2(\xi)-1)^2 d\xi
- {1\over 4} \int_r^\infty \xi (f^2(\xi)-1)^2 d\xi \sim -{\gamma_1\over r}
\ \ \ {\rm as} \ \ \ r\to\infty
\ee
where
\be
\gamma_1 = {1\over 4} \int_0^\infty \xi^2 (f^2(\xi)-1)^2 d\xi \simeq 0.675
\label{gamma1}
\ee
Thus $\epsilon\phi_1$ remains safely of order $\epsilon$, and the
linearized approximation is consistent, $A$ and $B$ taking the 
Barriola-Vilenkin (BV) form
\cite{BV}.

\subsection{Massive dilatonic gravity}

Now consider the case of a massive dilaton. We 
use $V(\phi)=2m^2\phi^2$ where the mass $m$ is measured in units 
of the Higgs mass. Obviously we do not expect this to be
the exact form of the potential, however, for small perturbations
of the dilaton away from its vacuum value, we might expect a quadratic
form to be a good approximation. Naturally we will have to check the
self consistency of such an approach. We expect 
$10^{-11} \le m \le 1$, representing a range for the unknown dilaton mass 
of 1 TeV - $10^{15}$ GeV. The dilaton equation is
\begin{eqnarray}
& & \left ( \frac{B'}{2AB}- \frac{A'}{2A^2} + \frac{2}{Ar} 
\right )\phi' + \frac{\phi''}{A} - m^2\phi = \nonumber \\
& & \epsilon \left ( (a+1)e^{2(a+1)\phi} \left ( \frac{f'^2}{2A}+ 
\frac{f^2}{r^2} \right ) + \frac{a+2}{4}e^{2(a+2)\phi} (f^2-1)^2 
\right ) \label{massivedilaton} 
\end{eqnarray}
Again we expand about flat space
\begin{eqnarray}
& & A = 1+ \epsilon A_1 + \ldots  \nonumber \\
& & B= 1+ \epsilon B_1 + \ldots \\
& & \phi= \epsilon \phi_1 + \ldots  \nonumber
\end{eqnarray}
To ${\mathcal{O}}(\epsilon)$, 
(\ref{massivedilaton}) gives
\be
\phi_1'' + \frac{2}{r}\phi_1' - m^2\phi_1= (a+1)\left ( 
\frac{f'^2}{2} + \frac{f^2}{r^2} \right ) + \frac{a+2}{4}(f^2-1)^2
\ee
which can be integrated to give
\begin{eqnarray}
&  \phi_1 = -\frac{1}{mr} e^{-mr} \int_0^r \xi \sinh m\xi 
\left ( (a+1)\left ( \frac{f'^2}{2} + \frac{f^2}{\xi^2} \right ) 
+ \frac{a+2}{4}(f^2-1)^2 \right ) d\xi & \nonumber \\
&  - \frac{1}{mr} \sinh mr \int_r^{\infty} \xi e^{-m\xi} \left ( 
(a+1)\left ( \frac{f'^2}{2} + \frac{f^2}{\xi^2} \right ) + 
\frac{a+2}{4}(f^2-1)^2 \right ) d\xi & 
\end{eqnarray}
Outside the core we find the leading order behaviour of $\phi_1$ to be
\be
\phi_1 \simeq -{(a+1)\over m} \int_0^r {e^{-mu}\over r^2-u^2}du
\simeq - {(a+1)\over m^2r^2}
\ee
for $a\neq -1$, and $\phi_1 = {\cal O} (1/(m^2r^4))$ for $a = -1$.
$A$ and $B$ will once again take their BV forms. 
Thus in contrast to the local cosmic string, for all values of $a$ there
is a diffuse dilaton cloud.

\begin{figure}
\begin{center}
\epsfig{figure=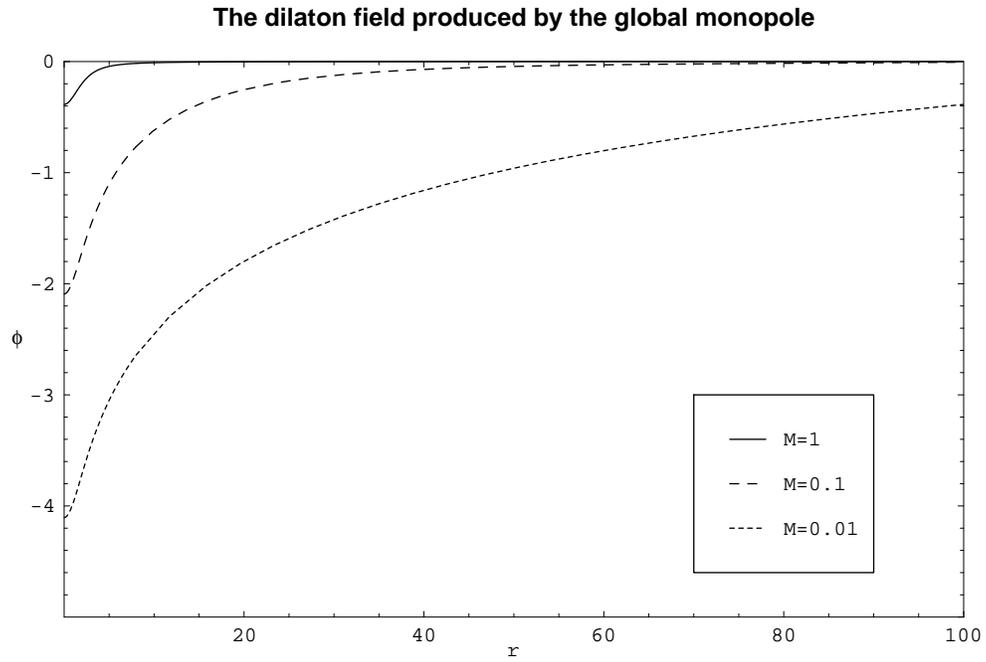,height=18cm}
\caption{The dilaton field surrounding a global monopole. The factor
of $(a+1)\epsilon$ has been factored out of the dilaton. Note the
contrast in the logarithmic dependence of the amplitude compared to
the reciprocal dependence of the fall off on the mass of the dilaton.
\label{fig:gdil}}
\end{center}
\end{figure}

\begin{figure}
\begin{center}
\epsfig{figure=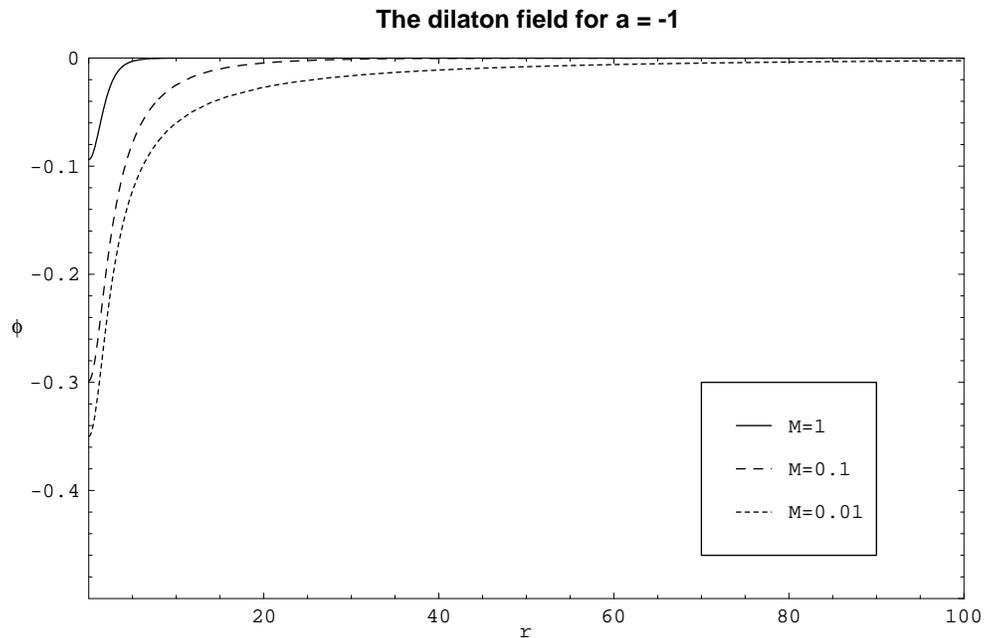,height=18cm}
\caption{The dilaton field for $a = -1$. Note that while the horizontal
scale is the same as for  figure \ref{fig:gdil}, the vertical scale is an order
of magnitude less. The effect of changing $m$ is much less pronounced, and
one sees the amplitude approaching a limit of approximately -0.35.
\label{fig:mdil}}
\end{center}
\end{figure}

We must now check that the dilaton remains small for all values of
$m$ under consideration. This is not only to verify the consistency of
our linearization prescription for solving the equations of motion,
but also to justify taking a quadratic approximation to the dilaton
potential. To see this, note that 
\begin{eqnarray}
\phi_1(0) &=& - \int_0^\infty \xi e^{-m\xi} \left ( 
(a+1)\left ( \frac{f'^2}{2} + \frac{f^2}{\xi^2} \right ) + 
\frac{a+2}{4}(f^2-1)^2 \right ) d\xi \nonumber \\
&=& -\gamma_2 - (a+1) \int_1^\infty {e^{-m\xi}\over\xi}d\xi
\end{eqnarray}
where
\be
\gamma_2 = \int_0^\infty \xi e^{-m\xi} \left ( (a+1) {f'^2\over2} +
{a+2\over4} (f^2-1)^2 \right ) d\xi + 
(a+1) \int_0^1 {f^2 e^{-m\xi}\over\xi} d\xi \simeq 1
\ee
is approximately independent of $m$\footnote{We find for $a = 0$, that
$\gamma_2 = 0.3$ for $m=1$, $0.9$ for $m=0.1$, and $1.1$ for
$m\leq 0.01$ to 2 significant figures.}, and the core of the monopole is taken
to be 1 for convenience. But,
\be
\int_1^\infty {e^{-m\xi}\over\xi} d\xi = \int_m^\infty {e^{-u}\over u} du
\sim - \ln m + ....
\ee
hence $\phi_1(0) = {\cal O} \left ( (a+1) \ln m\right)$, (for $a = -1$,
$\phi_1(0) \simeq 0.35$) and we can 
therefore
justify the approximation of taking a quadratic potential for $\phi$ (as 
well as the linearized approximation) provided $|\ln m| \ll \epsilon^{-1}$,
which is easily satisfied by the parameter ranges under consideration.
A plot of the dilaton field for various values of $m$ and
$a\neq -1$ is shown in figure \ref{fig:gdil}.
A plot of the dilaton field for $a = -1$ is shown in figure \ref{fig:mdil}.

\section{Discussion}

We have derived the metric and dilaton field of a global monopole
in low energy string gravity for an arbitrary coupling of the monopole
lagrangian to the dilaton : $e^{2a\phi} {\cal L}$. For the
massless dilaton, this modification generically destroys the 
good global behaviour of the monopole, making it singular. This is
because the dilaton multiplies the energy density of the monopole
worsening the already strong gravitational effect.
For $a = -1$, the metric is nonsingular and of the BV form in the
Einstein frame, the dilaton taking the asymptotic form
$\phi \simeq - \epsilon \gamma_1/r$, where $\gamma_1$ is given by
(\ref{gamma1}). In the string frame, the metric is given asymptotically
by
\be
\hat{ds}^2 = \left ( 1 - {2(M+\epsilon\gamma_1)\over{\hat r}} \right )
d{\hat t}^2 - \left ( 1 - {2(M+\epsilon\gamma_1)\over{\hat r}} \right )^{-1}
d{\hat r}^2 - (1 - \epsilon) {\hat r}^2 d\Omega^2_{\rm II}
\ee
with respect to suitably rescaled coordinates $\hat{t}, \hat{r}$.
Inputting our values for $M$ and $\gamma_1$, we see that in the string
frame the ADM mass of the global monopole is $\hat{M} \simeq - 0.055 
\epsilon$ which is substantially smaller than the Einstein ADM mass, but still
negative.

For the massive dilaton, to leading order
in the Einstein frame the metric takes the BV form.
The dilaton asymptotes $-(a+1)\epsilon/(m^2r^2)$
for $a\neq -1$, and $-\epsilon /(m^2r^4)$ for $a =-1$.
This power law fall-off of a massive scalar field seems
counterintuitive until one remembers that the dilaton is
in fact part of the gravitational sector of the theory,
and therefore couples to the energy momentum of the global monopole.
The slow fall off of this energy momentum is what supports the rather
diffuse dilaton cloud. We therefore have
a rather different, nebulous, dilaton cloud surrounding the global monopole as
opposed to the well defined cloud surrounding a local cosmic 
string\cite{GS}. To leading order in $1/r$, the asymptotic metric
in both string and Einstein frames is the same, and is identical to the
BV result.

Our results indicate that astrophysical bounds\cite{H,ABD} 
on global monopoles obtained
from their gravitational or metric field will be little altered by
the dilaton, except if we approach close enough to feel its
gravitational mass. On the cosmological scales therefore,
the gravity of global monopoles is unchanged. 
One might wish to ask how the dilaton effects monopole bounds more
directly, following the lines put forward by Damour and Vilenkin
\cite{DV}. Here we note that for $a \neq -1$, the dilaton fall-off is
as a power law in $mr$, therefore determined by a scale $m^{-1}$. 
We therefore expect the Damour-Vilenkin bound to hold, and such
global monopoles will be inconsistent with a low (TeV) mass dilaton.
For $a = -1$, the dilaton falls off as $(mr^2)^{-2}$, hence determined
by $m^{-1/2}$ (or $m^{-1/2}(\sqrt{\lambda}\eta)^{1/2}$ before
rescaling). We should therefore subsitute $m_\phi^{1/2}m_{\rm Higgs}^{1/2}$
for $m_\phi$ in the Damour-Vilenkin bound. This has the effect of weakening
the bound. For example, for a TeV mass dilaton, Damour and Vilenkin
quote a bound on the symmetry breaking scale of $\eta \leq 10^{13}$GeV
for the global monopole \cite{DV}. For $a = -1$, this is weakened to 
$\eta \leq 10^{14}$ GeV. Although this is obviously a weaker bound, 
it is in contrast to the local cosmic string where the bound would appear
to be removed for $ a = -1$.

In conclusion, the gravitational field of a global monopole in the
presence of a massive dilaton is broadly similar to that in
Einstein gravity, however, the presence of a diffuse dilaton cloud
leads to bounds on the energy scale of the global monopole
due to its dilaton production\cite{DV}. In contrast, for a massless dilaton 
the spacetime is only regular if $a = -1$, and even then
the ADM mass is significantly different from Einstein gravity.

\section*{Acknowledgements}

We would like to thank Filipe Bonjour and Caroline Santos for
helpful discussions. O.D.\ is supported by a PPARC studentship,
and R.G.\ by the Royal Society.

\end{document}